\documentclass[nologo,url,11pt,a4paper]{ETHpaper}

\usepackage{graphicx}                  
\usepackage[english]{babel}            
\usepackage{lscape}
\usepackage{rotating}
\usepackage[numbers]{natbib}
\usepackage{amsmath, amsthm, amssymb, amsfonts}
\usepackage{setspace}
\usepackage{epstopdf}
\usepackage{tikz}
\usepackage{slashbox}

\long\def\symbolfootnote[#1]#2{\begingroup%
\def\thefootnote{\fnsymbol{footnote}}\footnote[#1]{#2}\endgroup}

\title{The Community Structure of the Global Corporate Network}

\author{Stefania Vitali$^{1,2}$, Stefano Battiston$^{1}$\symbolfootnote[1]{Corresponding Author}\symbolfootnote[2]{SB acknowledges support from FET Open Project FOC "Forecasting Financial Crises" nr. 255987. }}
  \authoralternative{Stefania Vitali, Stefano Battiston}
\address{$^{1}$ Chair of Systems Design, ETH Zurich, Kreuzplatz 5, 8032
  Zurich, Switzerland\\
  $^{2}$ Dipartimento di Fisica e Chimica, Universit\`a degli Studi di Palermo, Palermo, Italy\\
  \url{stefania.vitali@unipa.it,sbattiston@ethz.ch}}

\begin{document}
\maketitle

\begin{abstract} 
We investigate the community structure of the global ownership network of transnational corporations. We find a pronounced organization in communities that cannot be explained by randomness. Despite the global character of this network, communities reflect first of all the geographical location of firms, while the industrial sector plays only a marginal role. We also analyze the network in which the nodes are the communities and the links are obtained by aggregating the links among firms belonging to pairs of communities. We analyze the network centrality of the top 50 communities and we provide the first quantitative assessment of the financial sector role in connecting the global economy. 
\end{abstract}

 \medskip
\medskip


\medskip
\medskip
\medskip

\section{Introduction}
\label{sec:motivation}

A recent work has studied the global structure of ownership network with respect to the issue of the corporate control \cite{vitali.ea11a}. In this paper, instead, we carry out an in-depth community analysis \citep{fortunato10} of the same network, in order to address questions concerning the level of geographical integration and the role of the financial sector in the global economy.


An economic network is a structure in which some economic entities, represented as nodes, are connected to some other entities by means of one or more specific types of relationships. Previous empirical studies in socio-economic networks, that are relevant for our work, include those focusing on: international trade \citep{garlaschelli.ea05a,fagiolo.ea09}, international financial exposures \citep{chinazzi2012post,minou2012global}, financial networks \cite{garlaschelli.ea05,bonanno2003topology,battiston.ea12,kaushik.ea12,boss.ea04,cont2011network,cajueiro2008role}
More in detail, regarding corporate governance, there are two main groups of works: (i) those on corporate boards, e.g., interlocking directorates \citep{battiston.ea03,conyon.ea06,davis.ea97}, and those on firm ownership \citep{corrado.ea06, vitali.ea11a, vitali.ea11b, glattfelder.ea09, kogut.ea01}. However, in these disciplines little attention has been devoted to the community structure. Apart from the study of \citep{piccardi.ea10}, which analyzes the community structure of two small networks, the Italian board and ownership networks, the other community analyses have focused on correlation networks in stock markets \cite{song.ea11} and in foreign exchange markets \cite{fenn.ea08}.

The study of the transnational corporation (TNC) ownership network reveals that corporations are well connected, with the large majority of the nodes belonging to a large connected component, which is itself organized in several communities. The community analysis is performed by applying the method of \citep{blondel.ea08} that belongs to the optimization class and is one of the few algorithms suitable for the investigation of large networks with no imposed constraints on the number of communities \citep{newman04}. In order to asses the robustness of the resulting community partitions, we have compared the community structure in the empirical network with the one obtained from a random link formation process, accounting for the constraints on the degree distribution and on the ownership structure \citep{maslov.ea02,zlatic.ea09,vitali.ea11b}. The comparison reveals that for the rewired networks the community structure is quite homogeneous across realizations and it differs considerably from the one of the empirical network, meaning that the community structure cannot be considered the result of a random pattern of link formation.  Furthermore, the analysis of the geographical and sector properties of the communities reveals that, while both features characterize the communities, the country dominance tends to be more pronounced than the sector dominance.  Finally, we consider communities as themselves forming a network, in which the link between any two given communities reflects the number of ownership relations among firms from the two communities. We assess the importance of each community in the network by using DebtRank, a centrality measure recently introduced in the complex systems literature in the context of economic networks \cite{battiston.ea12}. In particular, we apply this method to investigate the role played by the financial sector as a source of the connections among communities. We find that the community centrality changes drastically when we exclude from the sample the firms belonging to the financial sector. The difference in centrality quantifies the role played by the financial sector in the strength of the links among communities and, thus, in determining the potential impact that each community has on the others.

The paper is organized as follows: in Section\ \ref{sec:data} we present the database, the cleaning procedure we have performed, and we provide the basic network notions. In Section\ \ref{sec: community analysis} we introduce the community analysis and the community network is also studied. Section\ \ref{sec:role_fin_sector} investigates the role played by the financial sector in the ownership community network and, finally, Section\ \ref{sec:conclusion} concludes.

\section{Data}
\label{sec:data}



The dataset we investigate in this paper is the same that was analyzed in\ \cite{vitali.ea11a} and extracted, by means of the procedure explained in the following, from the Orbis 2007 database\ \footnote{URL: \texttt{http://www.bvdep.com/en/ORBIS} \label{foot}} containing information as of the last quarter of 2007 for more than 30 million economic actors (firms and shareholders). It includes the name of firms, their geographical localization (country and city), industrial classification (NACE) and several financial data. Moreover, the database includes data on about 12 million ownership relations, with information on the name of the shareholder and the amount of shares.

The procedure of extraction is an important part of the methods and it works as follows. We first identify all the transnational corporations (TNC), defined as those companies that are headquartered in one country and operate in at least one foreign country, by owning partially, at least 10\% of the shares, or wholly other companies\ \citep{OECD00}. We obtain a list of 43060 TNCs, located in 115 different countries. The major part of these TNCs have their headquarters in Europe and the US. Nevertheless, some of them are also located in off-shore countries like Bermuda (with 139 companies) and Cayman Islands (with 40 companies). Then, we explore recursively the neighborhood of the TNC companies in the whole database. Two recursive searches are applied: (i) we proceed downstream by identifying all the participated companies directly and indirectly owned by the TNCs with a breadth-first search procedure; (ii) we proceed upstream with the same procedure in order to find the shareholders that have direct and indirect paths leading to the TNCs. In this way, we assemble a network in which each node is connected to at least one TNC. 
The resulting network consists of 600508 nodes corresponding to economic entities and 1006987 links corresponding to corporate ownership relations.

In an ownership network, the nodes correspond to economic entities (e.g., companies or people owning equity shares) and the links to ownership relationships connecting them. We recall that a network $G$ is defined as the set of nodes $N$ and the set $E$ of edges represented by ordered pairs of nodes $\{ s, t \}$, with $s$ being the source and $t$ the target nodes of the edge $s \rightarrow t$. The weighted adjacency matrix of the network is $A=\{a_{i,j}\}$, where $a_{i,j}$ is the share that $i$ owns in $j$.  The network does not contain self-loops and the sum of shares of a firm held by other entities can not exceed 100\%, i.e., $\sum a_{ij} \le 1$, $\forall$ $i,j$. The in-degree, $k^{in}_i$, is the number of incoming links to a node $i$, that is, according to the convention we follow here, the number of shareholders. The out-degree, $k^{out}_i$, refers, instead, to the number of node $i$'s outgoing links and represents the number of firms in $i$'s portfolio. A connected component is a subgraph in which all the nodes can reach all the other nodes via an undirected path.


\section{Community Analysis}
\label{sec: community analysis}

\subsection{Community Detection Procedure}
\label{sec:methods}
In the field of complex networks, the notion of ``community'' corresponds, loosely speaking, to a subset of nodes that are more densely connected among themselves than with the nodes outside the subset. Several definitions of community and methods to detect communities have been proposed in the literature \citep[see][for a review]{fortunato10}.  Most algorithms can be distinguished in divisive \citep{newman.ea04}, agglomerative \citep{pons.ea06} and optimization-based \citep{newman06}. In the latter case, the goodness of the partitions is commonly assessed in terms of the so-called ``modularity'' \citep{newman.ea04}, defined as follows:
\begin{equation}
  \label{eq:1}
  Q = {1 \over {2l}} \sum_{i,j \in C} [a_{i,j} - {{k_ik_j} \over {2l}}],
\end{equation}
where $l$ is the total number of links and $k$ is the total degree ($k=k^{in}+k^{out}$). The modularity takes values between $-1$ and $1$ and compares the density of the links within the communities with those across communities. Positive values of modularity are a necessary but not sufficient condition for the presence of communities, since even random graphs can have positive values. Therefore, the values obtained have to be compared with those obtained in ensembles of rewired networks (see more below).

However, since the detection methods mentioned above suffer from certain shortcomings, in this study, we apply the method of \citep{blondel.ea08}, that belongs to the optimization class and is one of the few algorithms that, at the same time, are suitable: (i) to analyze large networks and (ii) to avoid \textit{ex-ante} assumptions on the number of communities to be detected or on their size \citep{newman04}. The algorithm works in successive stages. In the first stage, each node is assigned to a different community and, then, they are moved to the community of their neighbors if the new location increases the modularity. This process is applied to all nodes repeatedly and it ends when no more gains could be achieved. In the second and in each further stage, the communities previously found are treated as the node of a new network. The links between two communities are defined as the sum of the links between pairs of nodes belonging, each, to one of the two communities. For a detailed description of the algorithm, see \citep[see][]{blondel.ea08}. 

Further, we want to compare the community structure in the empirical network with the one resulting from a random link formation process. In order to account for constraints that arise from the degree distribution, it is custom to generate ensembles of synthetic networks by applying a degree-preserving rewiring procedure \citep{maslov.ea02,zlatic.ea09,vitali.ea11b}.
When rewiring the links in an ownership network, it is important to satisfy two main constraints: (i) the degree sequence, i.e., the number of outgoing and ingoing links of each economic entity; and (ii) the total number of ownership shares owned by the shareholders. 
We, thus, follow the procedure described in \cite{vitali.ea11b} (leaving out the additional constraint on the geographical location of firms and shareholders) to generate 20 realizations of synthetic rewired networks.

\subsection{Community Structure}
\label{sec:community-structure}
\begin{figure}[ht]
\centering
\includegraphics[scale=0.495]{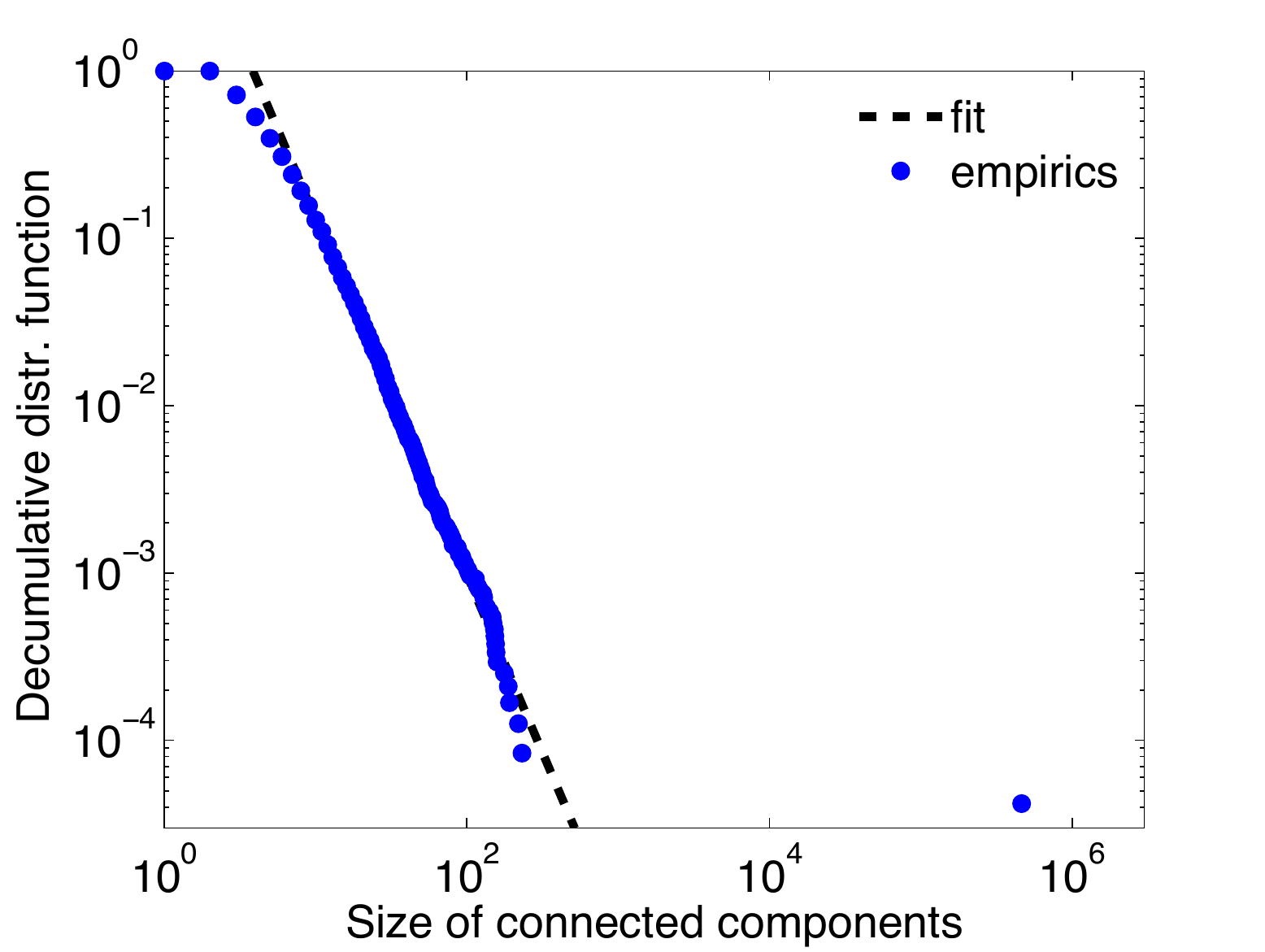}
\includegraphics[scale=0.4]{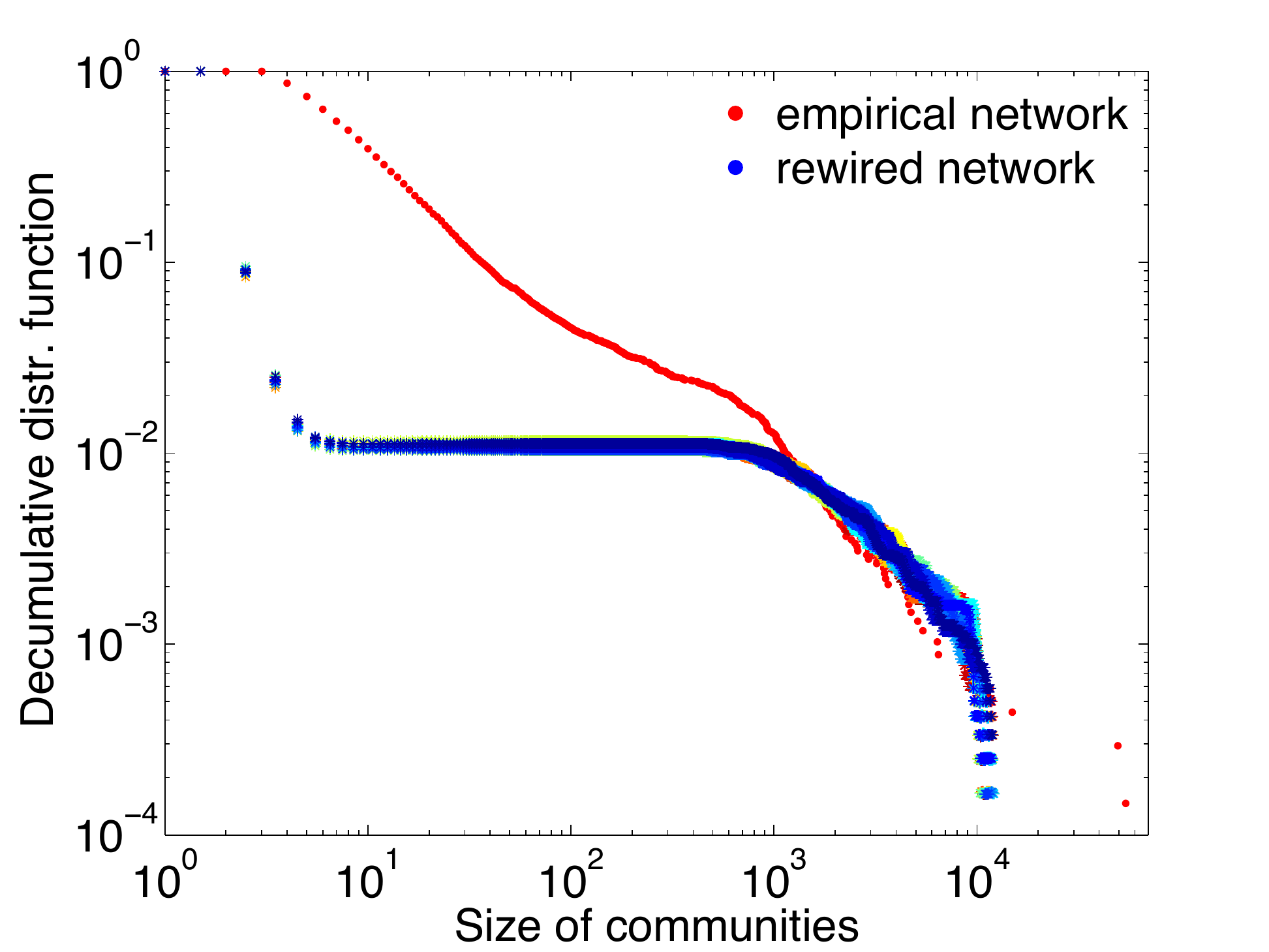}
\caption{(Left side) Decumulative distribution function of the size of the
  connected components. As a comparison, a power-law fit with exponent
  $−2.13$ is shown. (right side) Decumulative distribution function of the size of the
  communities in the empirical (in red) and reshuffled (in blue scale) networks.}
\label{fig:distr}
\end{figure}

The TNC ownership network is composed of 23825 connected components. The largest component contains 463006 nodes (77\% of the total), while the second largest connected component contains 230 nodes and 90\% of the components have less than 10 nodes. We report in Fig.\ \ref{fig:distr} (left side) the distribution of the component size. Notice how the data point corresponding to the largest component deviates from the trend of all the other components. A power law fit of the data points excluding the largest component is shown for reference, yielding an exponent of 2.13. 

In the following we have restricted our community analysis to the largest connected component (LCC). For the community analysis, we have utilized the unweighted version of the algorithm for the following reason. The weight of the links represent a share and, thus, a number between 0 and 1. The weight of edges pointing to big firms are typically small since no single shareholder is able to own or interested in owning a large fraction the capital. Therefore, assigning an importance to the links proportional to the share would result in treating shareholding relations to large firms as very weak links, inducing a bias in the communities towards small firms. A more appropriate way to proceed could be to try to account for the monetary value of the links, which depends on the value of the firm owned. However, this value is not available in the database for all the firms. Moreover, it would not be clear how to normalize the values in order to use them within the algorithm. 

In our empirical network, the algorithm stops the procedure of modularity
optimization after seven stages. In Tab.\ \ref{tab:community_size} we report for each stage: (i) the number of communities, (ii) the number of links, and (iii) the level of modularity. Notice that the number of communities decreases from 463006 (equal to the total number of nodes) to 6824. The number of links among communities decreases from 887515 to 25588, while the modularity increases from 0.0002 to 0.7344. Communities are very heterogeneous in size, ranging from those with only few nodes to the two main communities with about 50 thousands nodes. The $99\%$ of the communities contains less than 1000 nodes and $95\%$ less than 100. The distribution is reported in Fig.\ \ref{fig:distr}, right side, in red. 

\begin{table}  
\centering
\begin{tabular}{cccc}
\hline \hline
level & \# nodes & \# links & modularity\\ 
\hline
0 & 463006 & 887515 & 0.0002\\
1 & 82123 & 334368 & 0.5926\\
2 & 20021 & 93561 & 0.7076\\
3 & 9425 & 42321 & 0.7292\\
4 & 7226 & 28686 & 0.7339\\
5 & 6854 & 25896 & 0.7343\\
6 & 6824& 25588 & 0.7344\\
\hline \hline
\end{tabular}
\caption{Summary of the algorithm output of each community detection
  phase.}
  \label{tab:community_size}
\end{table}


For the rewired networks we find that they are quite homogeneous across realizations, in terms of their community structure, but they differ from the empirical network. They contain, on average, $11977\pm135$ communities, almost the double than the empirical network (6824).  The number of links among communities, however, is similar in the synthetic case ($28284\pm 654$) and in the empirical case ($25588$). 
In the empirical network, the value of modularity ($0.7344$) is about 30\% larger than in the rewired networks ($0.5242\pm 0.0002$). The difference is much larger than 3 times the std in the ensemble of rewired networks. We thus conclude that, in terms of modularity, the empirical network cannot be regarded as a realization of the ensemble of rewired networks. Similarly, the community size distribution in the empirical network (in red) strongly deviates from those obtained from the rewired networks (in blue color scale, in Fig.\ \ref{fig:distr}, right side). 


\subsection{Characterization of Communities in terms of Geography and Sector}
\label{sec:char-comm-terms}
In order to characterize the communities in terms of geography and sector, we start from the eight top largest communities, which together include $1/3$ of all the nodes. Tab.\ \ref{tab:communities_characterization} reports for each community: (1) the number of firms (\# firms); (2) the first and the second country by number of firms (C1 and C2); (3) the first and the second sector (Sector1 and Sector2) by frequency within that community. Moreover, we report the share of firms belonging to the leading country and sector (ShareC1 and ShareS1), as well as the country and sector concentration of each community as measured by the Herfindhal index (Herf. C and Herf. S). The share of firms located in the dominant country is rarely below 0.5 and the Herfindahl index is constantly above the limit between medium and high concentration (i.e., 0.25). However, for some communities the Herfindhal index also reveals that there is more than one dominating country. Indeed, for example in community 2, C1 and C2 contribute roughly equally to the community dominance. The role of geography is evident also in the first 100 biggest communities, which are almost all dominated by a single country located within the North America and Europe boundaries. The first Asian-dominated community is at rank $12^{th}$ by size\footnote{Asian firms are dominant only in few communities. The existence of only few and small Asian communities could be due to the traditional organization of Asian corporations in business groups \citep{laporta.ea99}, where members are densely connected in relatively small groups with few or no connections with external firms.}. For the characterization of communities in terms of sectors, we group all sectors in six macro-sectors (primary, manufacturing, services, financial intermediaries, real estates, renting and business activities, and state and social sectors). The share of the dominant sector in the top eight largest communities is generally smaller than the share of the dominant country, even if the number of possible macro-sectors, i.e. 6, is much smaller than the number of possible countries (Tab.\ \ref{tab:communities_characterization}). When we average across all communities with minimum size of 5 nodes, the share of firms belonging to the leading country of each community is 80\% in the empirical network as opposed to 25\% in the rewired networks. 
Similarly,  the share of firms belonging to the leading sector is 70\% against the 35\% in the rewired networks.

\begin{table}[t]
  \begin{center}
    \begin{tabular}{cccccccccc} \hline\hline {\em Comm} & {\em \# firms} &
     {\em Herf. C} & {\em ShareC1} & {\em C1} & {\em C2} & {\em Herf. S}& {\em ShareS1} & {\em S1} &
      {\em S2}\\ \hline
      1 & 54.065 & 0.362 & 0.588  & US  & CA  & 0.213 & 0.250 & services      & manufact.  \\
      2 & 49.475 & 0.208 & 0.428  & GB  & DE & 0.254 & 0.395 & business act. &services       \\
      3 & 14.917 & 0.578 & 0.756  & ES  & GB & 0.256 & 0.345 & business act. &services       \\
      4 & 11.658 & 0.669 & 0.816  & FR  & GB & 0.275 & 0.406 & business act. & services       \\
      5 & 10.475 & 0.685 & 0.825  & DE  & GB & 0.462 & 0.653 & business act. &financial int. \\
      6 &  6.462 & 0.539 & 0.730  & IT  & DE & 0.252 & 0.352 & services   &  business act. \\
      7 &  6.375 & 0.411 & 0.632  & DE  & GB & 0.312 & 0.446 & services      & business act.  \\
      8 &  5.420 & 0.265 & 0.421  & BE  & NL & 0.278 & 0.378 & business act. & financial int. \\
      \hline \hline
    \end{tabular}
  \end{center}
  \caption{Statistics on the top 8 communities. Herf. C is the
    Herfindhal index of the country concentration; ShareC1 is the share of
    companies localized in the country which dominate the community;
    C1 is the first dominant country; C2 is the second dominant country;
    Herf. S is the Herfindhal index of the sector concentration;
    ShareS1 is the share of companies active in the sector which
    dominate the community; S1 is the first dominant sector; S2 is the
    second dominant sector.}
  \label{tab:communities_characterization}
\end{table}

\begin{figure}[ht]
\centering
\includegraphics[scale=0.6]{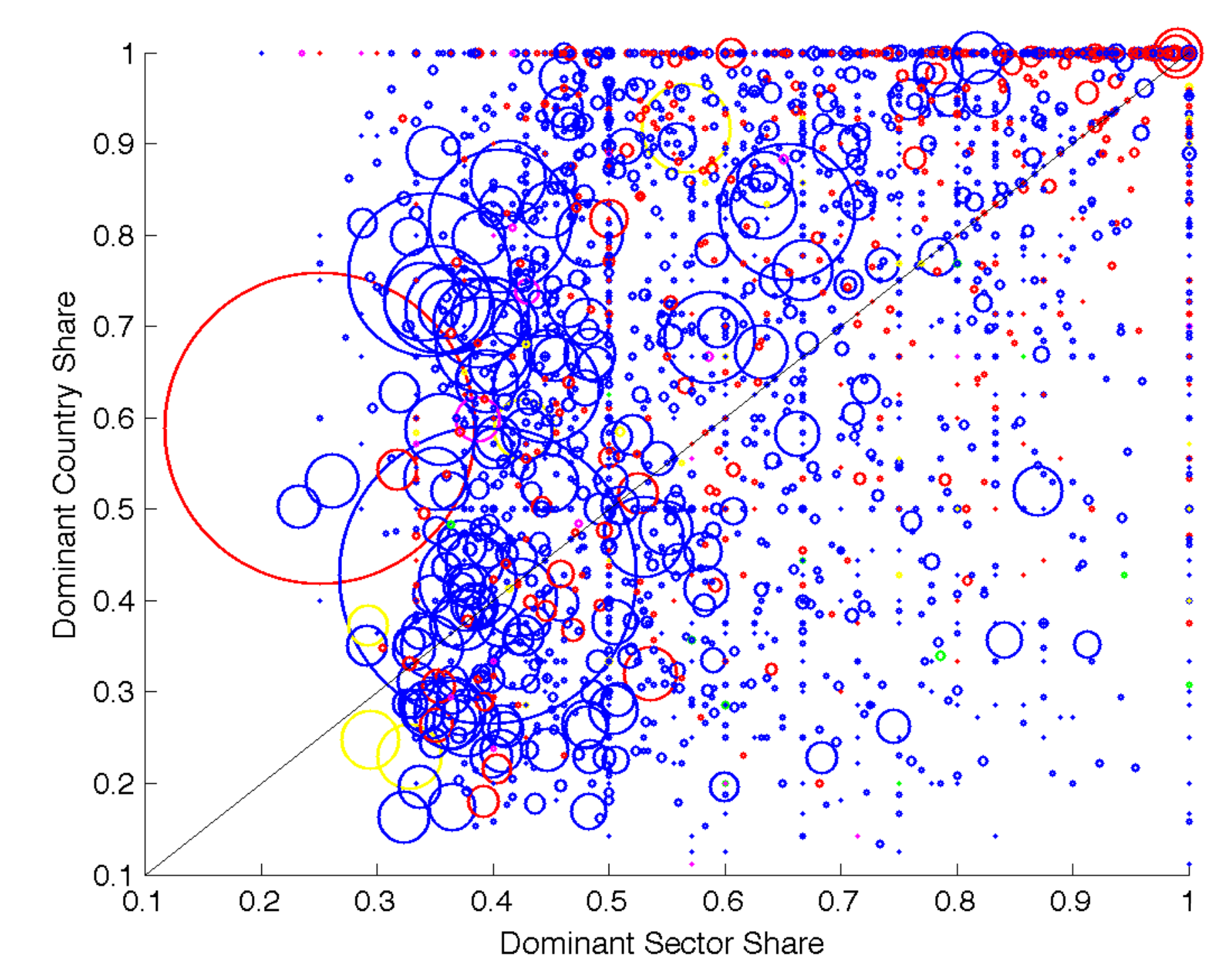}
\caption{Sector share versus country share of the dominant sector and
  country for each community. The size of the ball is proportional to the
  number of firms belonging to the community, while the color to the firm
  localization country (blue for EU, red for North American, yellow for Asian,
  green for fiscal paradise and magenta for all the other countries).}
\label{fig:scatter}
\end{figure}

Fig.\ \ref{fig:scatter} illustrates the geography and sector dominance of all the communities. The x and y axes represent the share of firms in the dominant sector and country of the community. The size of the circle reflects the size of the community in log scale and the color reflects the geographical region of the dominant country. As we can see, the fact that most circles are located above the diagonal implies that the country dominance tends to be more pronounced than the sector dominance. Many small communities are completely dominated by a country (value 1 or close to 1 on the y-axis).

Notice also that the two largest communities account together for about 1/5 of all the nodes and comprise companies mainly located in the US and Great Britain, respectively. Here below we provide some more details: 

\begin{itemize} 
\item The first biggest community includes 54065 economic entities. It is dominated by companies mainly located in North America (65\%), in particular in the US (59\%) and Canada (7\%), while 10\% of all the firms are located in three Asian countries (Japan, Taiwan and Korea). From a sector point of view, the nodes do not show a unique pattern: roughly 1/4 of the nodes belong, respectively, to the services,manufacturing and real estates, renting and business activities sectors. Finally, even if this community includes only 2283 TNCs (5\% of the total), in terms of operating revenue,it represents roughly 34\% of the total TNC value.  
\item The second largest community has 49475 members, of which 2004 TNCs accounting for the 17\% of the total operating revenue. Geographically speaking, the nodes belong, almost completely, to European countries (89\%), with Great Britain (42\%) leading the other countries (Germany is represented by 9.6\% of nodes, France by 6\%, Sweden by 5\% and Italy by 4\%.). The largest part of the companies are in the business activity industry (39\%), while the services and manufacturing sectors account for 20\% and 18\% respectively.  
\end{itemize}


Further, we apply the community characterization algorithm introduced in \cite{tumminello.ea11}. This statistical method reveals if a particular attribute of a community is ``over-expressed'', i.e., if its frequency in the community is larger than what expected from a random occurrence of the attribute across all the nodes in the network. On the one hand, by taking the location in a country as attribute and considering that in the whole network firms belong to 194 different countries, the algorithm finds that at least one country is over-expressed in all the large communities (i.e. larger than 250 nodes) and in about 50/\% of the smaller ones. In some communities, especially in the largest ones, more than one country is over-expressed (e.g., in the top two communities), see Tab.\ \ref{tab:communities_geo_over}. On the other hand, the sector attribute is less over-expressed than the geographical attribute. Indeed, only in roughly 30\% of all the communities at least one sector is over-expressed. Only the top ten largest communities display an over-expression of the sector, while many of the smaller community do not. The over-expressed sector for the top 8 largest communities is reported in Tab.\ \ref{tab:communities_sector_over}). These findings are replicated when we look also at smaller communities. Indeed, only the 0.28\% of the all geographical attributes of all the communities is over-estimated, about 65 times less than in the real community network, where about 20\% of the whole set of geographical attributes are over-expressed. The same result is worth for the sector attribute, with the 0.03\% of sectors over-expressed in the reshuffled communities, compared to the 15\% of the real ones.

Overall, the results of these analyses show that the community structure reflects the location of firms in the geographical space, while the role of the sector is much less important.

\begin{table}[t]
  \begin{center}
    \begin{tabular}{ccl} \hline\hline {\em Comm} & {\em \# firms} &
      {\em Countries over-expressed} \\ \hline 1 & 54.065 & US
      (30711/71343) JP (3562/4744) CA (3695/8830)
      AU(1095/2461) \\
      & & BM (299/577) ZA (407/998) IN (416/1180) KR
      (687/2719) KY (182/396) \\
      & & SG (318/1109) IL (117/219) BR (596/2746) HK (194/537) TW
      (987/5344) \\
      & & CN (264/923) ID (92/163) PH (72/167)
      TH (101/345)\\
      2 & 49.475 & GB (19235/112470) FI (1333/3690) SE (2047/10004) PL
      (940/3523) \\
      & & CH (809/4069) TH (157/345) GR (363/1874) NO
      (1126/7841) \\
      & & IT (1685/12899) RU (245/1361) IE (40/127) DK
      (911/7018) \\
      & & LU (299/2042) LI (13/31) IS (15/45)\\
      3 & 14.917 & ES (11052/21150) PT (297/3027) PE (76/320) AT
      (505/9694) \\
      & & BR (173/2746)\\
      4 & 11.658 & FR (9197/44854)\\
      5 & 10.475 & DE (8631/58138) LU (92/2042)\\
      6 &  6.462 & IT (4164/12899) RO (4/7)\\
      7 &  6.375 & DE (4004/58138) RU (161/1361) CH (262/4069)\\
      8 &  5.420 & NL (1424/27319) BE (2266/10651) LU (95/2042) \\
      \hline \hline
    \end{tabular}
  \end{center}
  \caption{Statistics on the community country characterization of the
    top 8 communities. The two numbers in parentheses are the number
    of occurrence of the country attribute in the selected community and in the
    whole network.}
  \label{tab:communities_geo_over}
\end{table}

\begin{table}[t]
  \begin{center}
    \begin{tabular}{ccl} \hline\hline {\em Comm} & {\em \# firms} &
      {\em Sectors over-expressed} \\ \hline 
      1 & 54.065 & financial interm. (8936/46632) manufacturing
      (11912/66502) \\
      && primary (1841/7010) services (12168/100445) \\
      2 & 49.475 & primary (940/7010) financial interm. (5339/46632) \\
      3 & 14.917 & services (4199/100445) state and social (860/21376)\\
      & & business activities (4467/130685)\\
      4 & 11.658 & business activities (4416/130685) manufacturing
      (2092/66502)\\
      5 & 10.475 & business activities (6156/130685) financial interm. (1409/46632)\\
      6 &  6.462 & services (1803/100445)\\
      7 &  6.375 & services (6375/100445) state and social (446/21376)
      primary (151/7010)\\
      8 &  5.420 & financial interm. (1533/46632) business activities (1932/130685)\\
      \hline \hline
    \end{tabular}
  \end{center}
  \caption{Statistics on the community sector characterization of the
    top 8 communities. The two numbers in parentheses are the number
    of occurrence of the sector attribute in the selected community and in the
    whole network.}
  \label{tab:communities_sector_over}
\end{table}

\subsection{Sub-communities: the inner structure of communities}
\label{sec:sub_communities}

\begin{figure}
 \centering
\includegraphics[width=12cm]{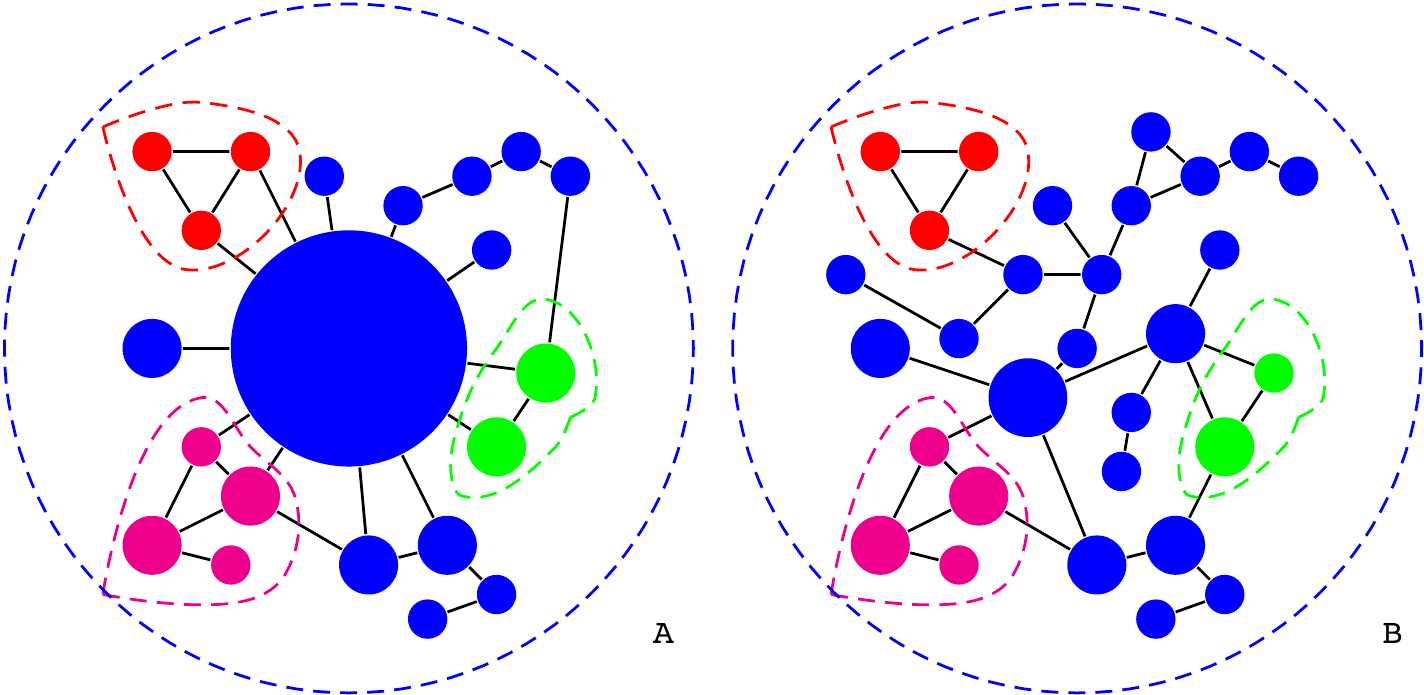}
\caption{Illustration of the two types of sub-community
  structures. One large and many small sub-communities, all
  geographically clustered (left side). The two biggest communities display this
  kind of structure. Many small sub-communities, geographically
  clustered. Most other communities show this type of organization (right side).}
\label{fig:sub_communities}
\end{figure}

In order to investigate the inner structure of a given community, we track by which sub-communities it is composed among those identified at stage 1 of the community detection algorithm and we, then, look at the statistics on the size. For instance, we find that the firms in the top largest sub-community belong also to the first top community and represent a major part of that community (i.e., 42121 firms out of 54065). Similarly the firms in the second largest sub-community belong also to the second top community (with 19247 firms out of 49475). Conversely, all the other sub-communities comprise less than 600 nodes. Tab.\ \ref{tab:subcommunities_stats} reports the number of sub-communities of each community along with the Herfindhal index computed on the size of such sub-communities. The Herfindhal index values for the first two communities (i.e., 0.607 and 0.151) indicate that in both cases the sub-communities are dominated in size by the largest sub-community. In contrast, from the third community onward, the Herfindhal index is at least 10 times smaller.  This suggests that the inner structure of the two top largest communities differs from the one of all the other communities. While the two largest communities have a star-like structure with one large sub-community acting as a hub, the other communities have a much less hierarchical structure (see Fig.\ \ref{fig:sub_communities}).

All sub-communities display a geographical characterization.
For instance, 70\% of the sub-communities of the largest community consist of firms located in only one country.


\begin{table}
  \begin{center}
    \begin{tabular}{cccc} \hline\hline {\em Comm} & {\em \# firms } &{\em \# subcomm} &
      {\em Herf. sub-comm. size} \\ \hline
      1 & 54.065 & 3997  & 0.607 \\
      2 & 49.475 & 7036  & 0.151      \\
      3 & 14.917 & 3007  & 0.001\\
      4 & 11.658 & 2641  & 0.001\\
      5 & 10.475 & 1816  & 0.004 \\
      6 &  6.462 & 1090  & 0.003 \\
      7 &  6.375 & 1426  & 0.002  \\
      8 &  5.420 & 921  & 0.016\\
      \hline \hline
    \end{tabular}
  \end{center}
  \caption{Statistics on the sub-communities of the top 8 communities.}
  \label{tab:subcommunities_stats}
\end{table}

\section{The Community network and the role of the financial sector}
\label{sec:role_fin_sector}

Communities can be regarded themselves as forming a network in which the link between any two given communities is the number of links between firms in the two communities. The network of the top eight communities by size is shown in Fig.\ \ref{fig:net_naces}, left side.

In order to gain further insight into the whole community network, we start with a basic network analysis. The bow-tie decomposition of the network yields to the largest strongly connected component (LSCC) of $528$ nodes, an IN-component of $309$ nodes and an OUT-component of $5987$. Notice that in the community network the LSCC relative size (7.74\%) is much larger than in the firm-shareholder network (1347 nodes, \cite{vitali.ea11a}). This has a simple explanation. Consider a community in the OUT. Since links among communities are obtained aggregating links between shareholders and firms, it is enough for that community to have one of its firms investing in a firm belonging to one of the communities in the LSCC to make the community enter the LSCC. The LSCC is still much smaller in relative terms than in other paradigmatic real-world networks, such as wikipedia or the web \cite{broder.ea00}. The degree statistics yields: average in-degree $=2.5 \pm 8.7$, max in-degree $= 557$,  average out-degree $= 2.5 \pm 62$, max out-degree $= 4146$. For the shortest paths we find: maximum $=4$; average $2.43\pm 0.51$. The link density is $3.1 * 10^{-4}$.


Financial intermediaries are well integrated in the network and hold many ownership shares in companies belonging to both the non-financial and the financial sector. In fact, in our sample although they represent only a small fraction (9\%), at the same time they account for the 36\% of all the ownership relations in the network. Many of these relations appear to have a strategic nature. Indeed, the financial intermediaries hold shares larger than 5\% in 13\% of non-financial companies and in 60\% of other financial companies. 
In order to assess the role of the financial sector as a source of the connections among communities, we repeat the above statistics for the network of communities obtained by removing from each community the firms (46632 in total) that belong to the financial sector (i.e., companies having NACE codes in the classes [6500-7000) and named ``financial intermediation, except insurance and pension funding''), and all their links (351587 in total).

The bow-tie decomposition yields now $903$ isolated nodes. The LSCC (381 nodes) shrinks by 25\%, the IN-component (226 nodes) by 27\% and the OUT-component (4799 nodes) by 20\%. The degree statistics yields: average in-degree $=2.3 \pm 6.5$, max in-degree $= 394$, average out-degree $= 2.3 \pm 49$, max out-degree $= 3157$. For the shortest paths: maximum $=4$; average $2.45\pm 0.52$. The link density is $3.6 * 10^{-4}$. Because some links have been removed, the degree statistics has to decrease and the shortest path statistics has to increase, but the change is small.  

As we notice, overall the topology remains close to the case with the financial sector. In contrast, the removal of the financial sector has a strong impact at the level of the weight of the links among communities. Indeed, after removing the firms in financial sector, the number of ownership relations among firms decreases sharply (see Tab.\ \ref{tab:communities_nofinance}). For instance, in the $8^{th}$ community (the one with the highest share of financial intermediaries), the number of firms decreases by 29\% and the number of direct links by 42\%. On the other hand, the community less affected by the financial sector is the $7^{th}$, with 3\% of removed companies and a reduction of only 340 links. Overall, the links among communities decrease more than proportionally w.r.t. the internal links. As shown in Tab.\ \ref{tab:network_communities_nofinance}, apart from the $7^{th}$ biggest community which counts a small number of financial intermediaries, such result holds for all the communities analyzed. In some cases, the decrease is very strong. For example, the $1^{st}$ biggest community experiences a drop of 2/3 of the links within itself, and of 75\% of the links directed to the $2^{nd}$ biggest community. 


\begin{table}[h]
\centering
\begin{tabular}{c r r rr r r}
\hline \hline
&\multicolumn{3}{c}{With financial sector} &
\multicolumn{3}{c}{Without financial sector} \\
{\em community}&{\em \# firms}&{\em \# rel.}&{\em density}&{\em \# firms}&{\em \# rel.}&{\em density}\\
\hline
1 & 54.065 & 256.607 & 8.779e-05 & 45.129 & 78.040 & 3.847e-05\\
2 & 49.475 & 109.880 & 4.489e-06 & 44.136 & 52.713 & 2.712e-05\\
3 & 14.917 &  20.799 & 9.348e-05 & 13.529 & 15.726 & 8.892e-05\\
4 & 11.658 &  14.186 & 1.143e-04 & 10.487 & 7.545 & 8.336e-05\\
5 & 10.475 &  12.893 & 1.175e-04 & 9.066 & 6.627 & 1.015e-04\\
6 &  6.462 &   7.812 & 1.711e-04 & 5.781 & 6.541 & 1.530e-04\\
7 &  6.375 &   7.952 & 1.956e-04 & 6.208 & 8.051 & 1.847e-04\\
8 &  5.420 &   7.876 & 2.681e-04 & 3.887 & 3.824 & 1.871e-04\\
\hline \hline
\end{tabular}
\caption{Statistics on the largest communities with and without financial
  intermediaries. The size, the number of directed relations and the
  density are reported.}
\label{tab:communities_nofinance}
\end{table}

\begin{table}
  \centering
\footnotesize{
  \begin{tabular}{crrrrrrrr}
   \hline \hline

    \backslashbox{$Start$}{$End$}&{$1^{st}$}&{$2^{nd}$}&{$3^{rd}$}&{$4^{th}$}
    &{$5^{th}$}&{$6^{th}$}&{$7^{th}$}&{$8^{th}$}\\
    \hline
    {$1^{st}$} & 56.948 & 3.881 & 177 & 64 & 68 & 60 & 70 & 47\\
    & (256.607) & (29.747) & (810) & (363) & (306) & (260) & (225) & (256) \\
    {$2^{nd}$} & 1.818 & 41.327 & 218 & 102 & 91 & 119 & 184 & 73\\
    & (19.840) & (109.880) & (1.054) & (727) & (476) & (522) & (298) & (434) \\
    {$3^{rd}$} & 10 & 63 & 14.460 & 8 & 0 & 12 & 4 & 2\\
    & (41) & (351) & (20.799) & (16) & (3) & (17) & (4) & (4) \\
    {$4^{th}$} & 2 & 35 & 9 & 8.090 & 2 & 5 & 13 & 6\\
    & (14) & (236) & (20) & (14.186) & (5) & (11) & (13) & (23) \\
    {$5^{th}$}  & 4 & 31 & 2 & 1 & 7.050 & 0 & 15 & 2\\
    & (35) & (169) & (15) & (9) & (12.893) & (4) & (24) & (7) \\
    {$6^{th}$} & 6 & 37 & 13 & 2 & 1 & 4.640 & 9 & 1\\
    & (13) & (186) & (22) & (6) & (4) & (7.812) & (11) & (2) \\
    {$7^{th}$} & 6 & 15 & 5 & 4 & 18 & 10 & 7.193 & 0\\
    & (7) & (23) & (7) & (5) & (23) & (11) & (7.952) & (0) \\
    {$8^{th}$} & 1 & 9 & 1 & 1 & 0 & 1 & 4 & 3.202\\
    & (458) & (686) & (45) & (32) & (10) & (4) & (6) & (7.876) \\
   \hline \hline
    \end{tabular}
 }
    \caption{Network of the largest communities with (in parenthesis) and
      without financial intermediaries. ``Start'' refers to the communities from
      which the ownership relations depart, ``End'' to the
      target communities.}
  \label{tab:network_communities_nofinance}
\end{table}

Finally, in order to assess the relative importance of each community we use DebtRank, a centrality measure recently introduced in complex networks literature in the context of economic networks \cite{battiston.ea12}. Beyond the interpretation in terms of economic loss due the distress of one or more nodes in the network, DebtRank can be used as a measure of importance, once a network of impact is defined. Here, we define the impact of community $i$ over $j$ as the ratio between the number of investments of community $j$ into $i$ over the number of investments within community $j$= $W_{ij}= \beta \, A_{ji}/A_{jj}$, where $\beta$ is a rescaling factor that for visualization purposes we set equal to the number of nodes in the network under observation, $\beta= 50$.
Notice that we are not interested in carrying out a stress-test. We only aim to compare the importance of the communities in the case with and without the financial sector. Traditional measures of centrality are not well suited for this purpose. For instance, Eigenvector Centrality is defined only on strongly connected graphs, or equivalently on undirected graphs. Other measures of impact, e.g., \cite{kaushik.ea12}, require a normalization of the impact matrix which then prevents from making an absolute comparison of the importance of a given node across different networks (see \cite{battiston.ea12} for more details). 

\begin{figure}[h]
  \centering
\hspace{-1.5cm}\includegraphics[width=0.49\textwidth]{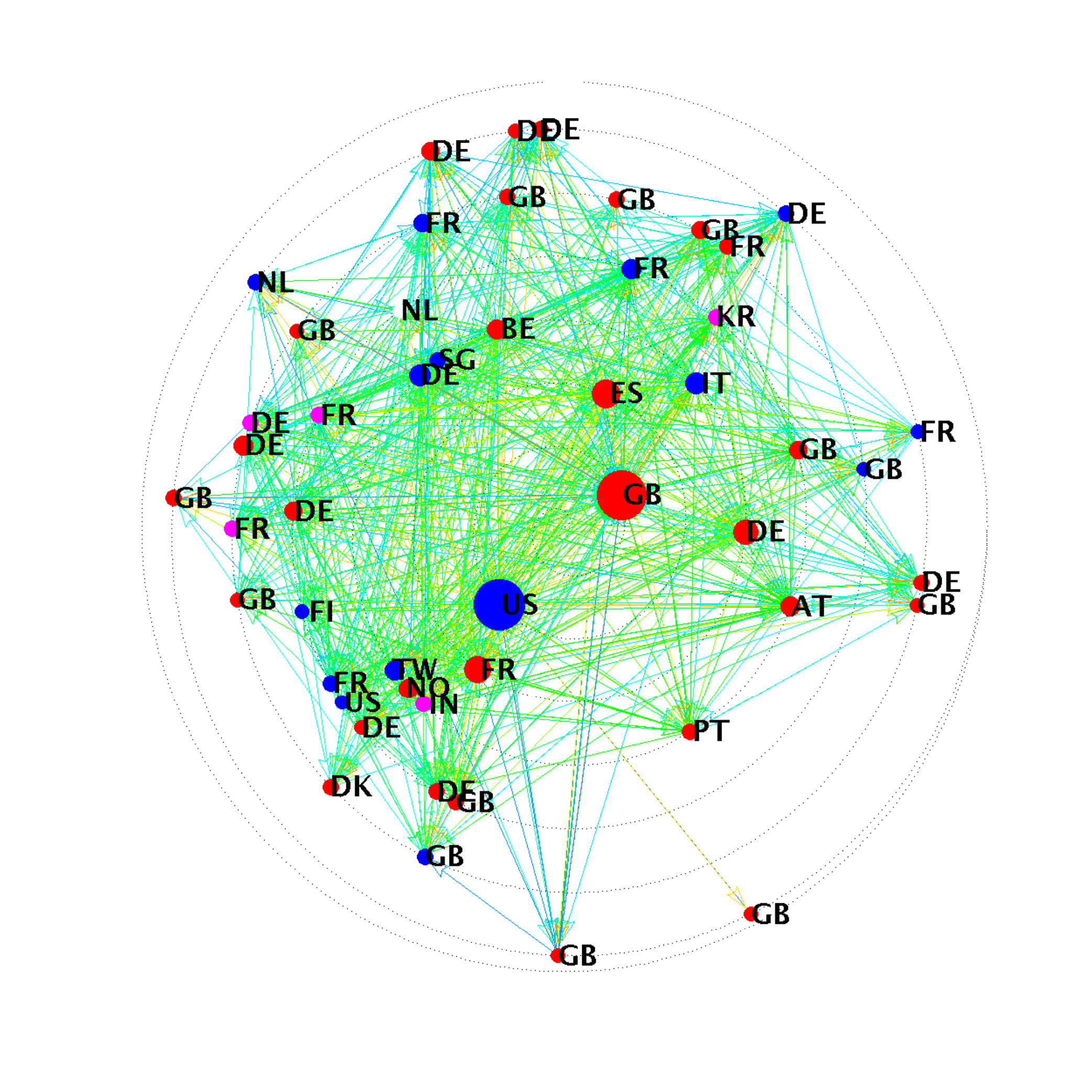}
\includegraphics[width=0.49\textwidth]{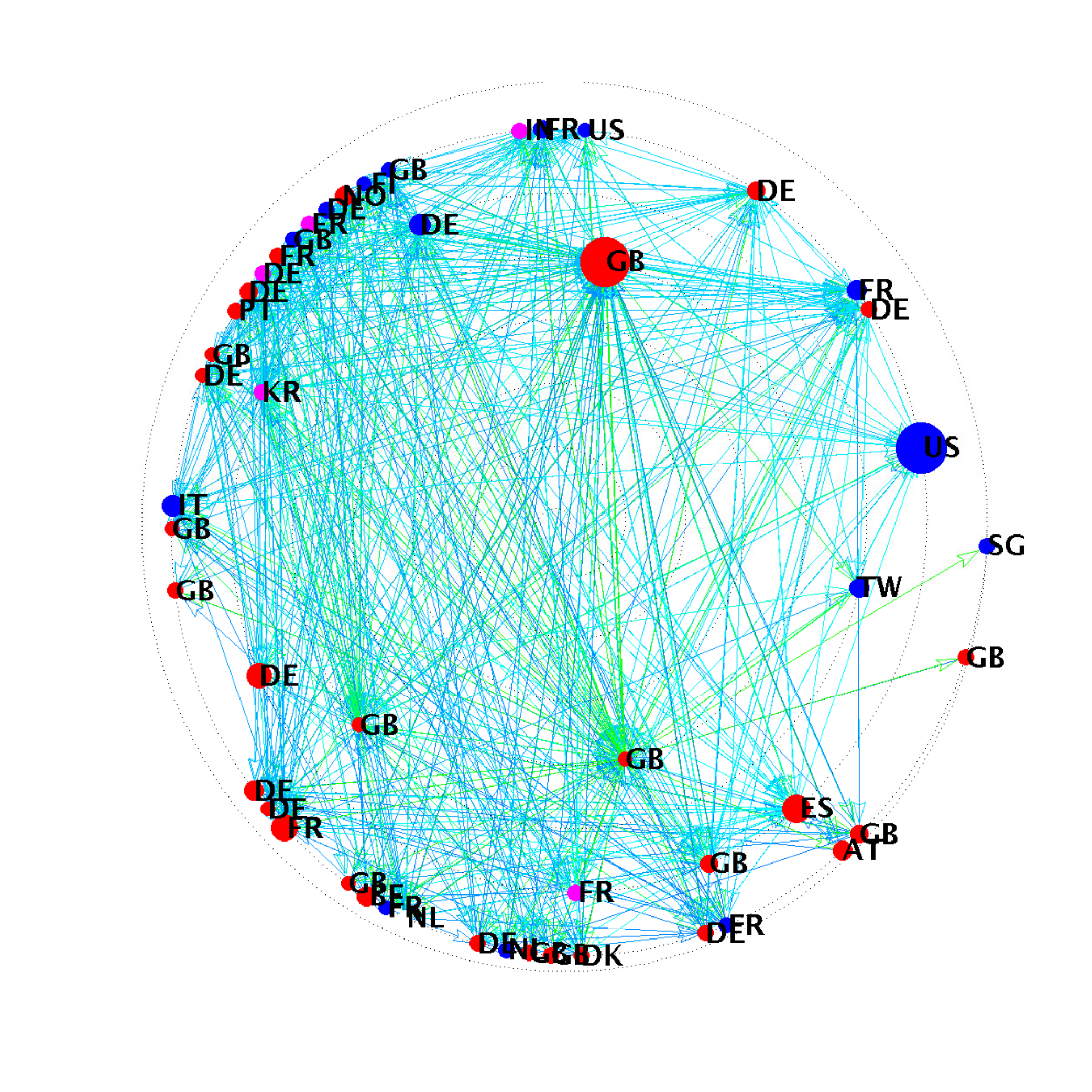}
\caption{Nodes represent the communities. Outgoing links represent the estimated potential impact of a community to another one. The closer a node is to the center the higher is its DebtRank (e.g., its centrality). The size of the node reflects the number of firms in the community (the size is set larger than a minimum, for visualization purposes). The color of the nodes corresponds to the dominant sector (red= business activities, blue = services, magenta= manufacturing), while the label indicates the dominant country. The color of a link reflects the DebtRank of the node from which it originates (see \cite{battiston.ea12} for more detail on the figure construction). Community network in the original dataset (left side) and after removing from the community partition the firms in the financial sector  (right side).}
\label{fig:communities_debtrank}
\end{figure}

Fig.\ \ref{fig:communities_debtrank}, left side, illustrates the network of the top 50 communities in a diagram where the position of each community reflects its centrality, as measured by DebtRank. More central communities are located in the center of the diagram. The size of each node is proportional to the number of firms in the community, the color corresponds to the dominant sector, while the label indicates the dominant country. As we can expect, the top communities by size are also more central. Fig.\ \ref{fig:communities_debtrank}, right side, illustrates the network of the top 50 communities after removing the firms in the financial sector. In this case, the top communities lose much of their centrality. As we can see, while the topological properties that do not account for the weight of the links are only moderately affected by the removal of the financial sector, the centrality computed with DebtRank, which do take weights into account, changes drastically. The difference in centrality quantifies the role played by the financial sector in the strength of the links among communities and, thus, in determining the potential impact that each community has on the others.


\begin{figure}[ht]
 \centering
\hspace{-2cm}\includegraphics[width=6cm]{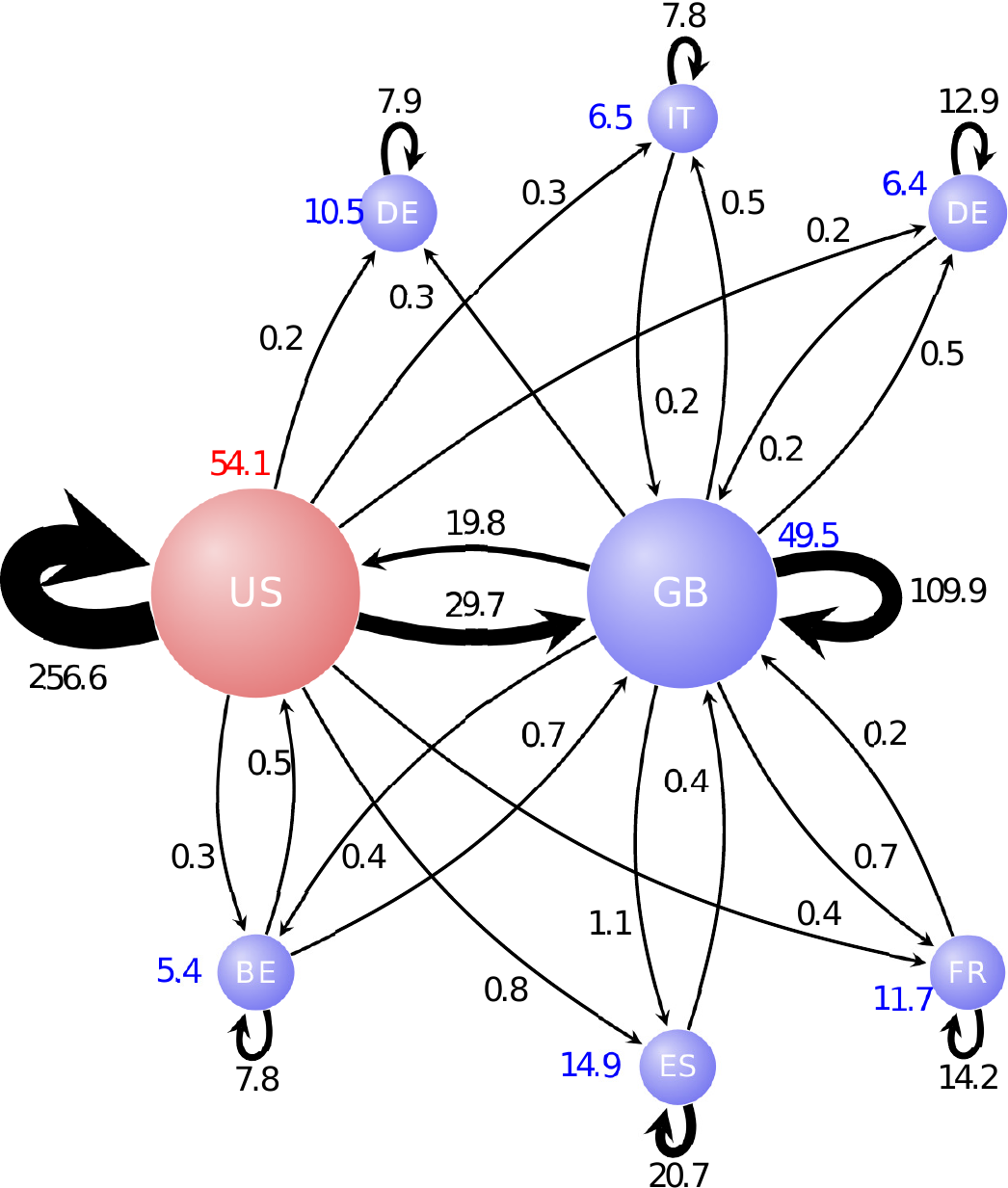}
\caption{Edge labels indicate the number of links between communities (in thousand). 
  The links with less than 50 ownership relations have been omitted. 
}
\label{fig:net_nace}
\end{figure}

\section{Conclusion}
\label{sec:conclusion}

In this paper, we have investigated the community structure of the global corporate network, i.e. the network of ownership among transnational corporations. This network is obtained from a large database of corporate information with a snowball procedure that starts from a list of about 43 thousand transnational corporations and recursively explores all the incoming and outgoing ownership relations. 

We have found that the organization in communities is pronounced and cannot be explained by randomness. Moreover, most communities are characterized by a dominant country, in the sense that the fraction of firms belonging to that country are not only the (relative) majority, but are over-expressed with respect to what would happen if the nationality is distributed at random among the firms. The characterization in terms of sectors is significant, but less pronounced than the one for countries. Thus, we conclude that the global corporate network is strongly clustered in communities, where geography is the major driver while sector is not so important.

We have also analysed the network in which nodes are the communities and links are obtained aggregating the links among the firms belonging to pairs of communities. In order to assess the role of the financial sector in the architecture of the global corporate network, we have analysed the centrality of the top 50 communities by means of the DebtRank algorithm. This has allowed us to obtain an absolute (as opposed to relative) measure of the importance of each community, which we have then used to compare the case with and without the firms in the financial sector. The difference between these two cases has provided a first quantitative assessment of the role of the financial sector in connecting the global economy. 


\clearpage \newpage

\bibliographystyle{unsrt}



\end{document}